\documentclass[12pt,a4paper]{article}

\usepackage{latexsym}
\usepackage{amsmath}
\usepackage{amsfonts}
\usepackage{amssymb}

\newcommand{\be}{\begin{equation}}
\newcommand{\ee}{\end{equation}}
\newcommand{\bea}{\begin{eqnarray}}
\newcommand{\eea}{\end{eqnarray}}

\newcommand{\bsq}{\boldsymbol{q}}
\newcommand{\bsp}{\boldsymbol{p}}
\newcommand{\qed}{{\emph{Q.E.D.}}}

\def\cG{{\mathfrak{g}}}                 %
\def\cO{{\cal O}}                       %
\def\cC{{\mathcal C}}                   %
\def\cH{{\mathcal H}}                   %
\def\ri{{\mathrm{i}}}                   %
\def\1{{\mbox{\boldmath $1$}}}          %
\def\ext{\mathrm{ext}}                  %
\def\red{\mathrm{red}}                  %
\def\bC{{\mathbb C}}                    %
\def\bN{{\mathbb N}}                    %
\def\bR{{\mathbb R}}                    %
\def\bZ{{\mathbb Z}}                    %
\def\tr{\mathrm{tr}}                    %
\def\diag{\mathrm{diag}}                %
\def\bX{{\boldsymbol{X}}}               %
\def\fu{{\mathfrak{u}}}                 %

\begin{document}

\vspace*{0.5cm}
\begin{center}
{\Large \bf Superintegrability of rational
Ruijsenaars-Schneider systems and their action-angle duals}
\end{center}

\vspace{0.2cm}

\begin{center}
V.~Ayadi${}^{a}$,
 L. Feh\'er${}^{a,b}$ and T.F. G\"orbe${}^{a}$\\

\bigskip

${}^a$Department of Theoretical Physics, University of Szeged\\
Tisza Lajos krt 84-86, H-6720 Szeged, Hungary\\

\bigskip

${}^{b}$Department of Theoretical Physics, WIGNER RCP, RMKI\\
H-1525 Budapest, P.O.B. 49,  Hungary \\

\bigskip

\end{center}

\vspace{0.2cm}

\begin{abstract}
We explain that the action-angle
duality between the rational Ruijsenaars-Schneider  and hyperbolic
Sutherland systems implies immediately the maximal superintegrability of
these  many-body systems.
We also present a new direct  proof
of the Darboux form of the reduced symplectic structure that
arises in the `Ruijsenaars gauge' of the symplectic reduction underlying this case of
action-angle duality.
The same arguments apply to the $BC_n$ generalization of
the pertinent dual pair, which was recently studied by Pusztai developing
a method utilized in our  direct calculation of the reduced symplectic structure.

\end{abstract}

\newpage

\section{Introduction}

The subject of superintegrability can be regarded as an offspring of the Kepler problem, which
is `more integrable' than motion in an arbitrary spherically symmetric potential
due to the existence of the extra conserved quantities provided by the Runge-Lenz
vector.
Recently we witnessed intense studies of superintegrable dynamical systems
motivated partly by interesting examples and partly by the natural goal
to classify systems with nice properties.
See, for example, \cite{BEHR,KKM,TW} and references therein.

Let us briefly recall the relevant notions of integrability for a Hamiltonian system $(M, \omega, H)$
living on a $2n$-dimensional symplectic manifold.
Such a system is called \emph{Liouville integrable} if there
exist $n$  independent functions $h_i \in C^\infty(M)$ ($i=1,\ldots,
n)$ that are in involution with respect to the Poisson bracket and
the Hamiltonian can be written as $H=\cH(h_1,...,h_n)$ through some smooth function $\cH$
of $n$ variables.
Importantly, one has to require also that
the flows of the $h_i$ are all complete.
A Liouville integrable system $(M,\omega,H)$ is termed
\emph{maximally superintegrable}  if it admits $(n-1)$ additional
constants of motion, say $f_j\in C^\infty(M)$, such that
$ h_1, \ldots, h_n, f_1, \ldots, f_{n-1}$
are functionally independent\footnote{Below the term superintegrable will always mean
maximally superintegrable.}.
The generic trajectories of $(M,\omega,H)$ are then given by the
connected components of the $1$-dimensional joint level surfaces
of the $(2n-1)$ constants of motion.
As a consequence, those trajectories of $(M,\omega,H)$ that stay inside
some  compact  submanifold of $M$ are necessarily
homeomorphic to the circle, since they are connected and compact 1-dimensional manifolds.
This implies that Liouville integrable systems having compact Liouville tori
are rarely superintegrable, because their trajectories are usually not closed.
On the other hand, it is common knowledge,  supported by
rigorous results \cite{GZ}, that systems describing repulsive interactions of particles
are superintegrable. Concretely, the scattering data provided by
the asymptotic particle momenta and differences of their conjugates
yield sufficiently many constants of motion. More abstractly \cite{Thirring},
the classical wave maps  furnish symplectomorphisms to obviously superintegrable free systems.

The aim of this contribution is to explain the superintegrability of
the celebrated rational Ruijsenaars-Schneider \cite{RS} and hyperbolic
Sutherland systems \cite{Sut,CalRag} in a self-contained manner.
Since these one-dimensional many-body systems support factorizable scattering \cite{SR88},
their superintegrability is not surprising.
However, we shall not use any scattering theory argument, which usually requires
non-trivial analysis of the dynamics. Instead of scattering theory, we shall directly rely on
special features of the `action-angle maps' of these Liouville integrable systems.
Indeed, it is known that these two systems form a dual pair in the sense that
they live on symplectomorphic  phase spaces, and the particle-positions
of each one of the two systems serve as action-variables of the other system.
The duality property was discovered by Ruijsenaars \cite{SR88} in his direct construction
of   `action-angle maps' that realize the introduction of action-angle variables.
More recently \cite{FK}, this duality has been fitted into the geometric
framework of symplectic reduction, which we shall utilize for showing superintegrability.

In Section 2, based on \cite{AF}, we recall
the elementary observation that Liouville integrable systems admitting
global action-angle maps of maximally non-compact type  are
maximally superintegrable.
Then, in Section 3, we explain how the geometric picture behind the rational Ruijsenaars-Schneider
and hyperbolic Sutherland systems permits to see easily that their
action-angle maps are the inverses of each other and are of maximally non-compact type.
In Section 4, we point out that this mechanism applies also to the
generalized  Ruijsenaars-Schneider and Sutherland systems that are associated with
the $BC_n$ root system.
The $BC_n$ generalization of the pertinent dual action-angle maps was recently
developed by Pusztai  \cite{P1,P2}.
In Appendix A, we take the opportunity
 to apply his ideas for
improving the previous (correct but not self-contained) calculation of the reduced
symplectic structure given in \cite{FK}.

\section{Action-angle maps of maximally non-compact type}

In scattering systems the canonical conjugates of the actions run over the line.
Later we shall exhibit interesting examples where the canonical transformation
to these Darboux variables represents  an action-angle map of maximally non-compact type as defined below.

Consider a Liouville integrable Hamiltonian system
$(M, \omega, H)$ possessing the $n$ Poisson commuting,  independent
constants of motion $h_i\in C^\infty(M)$, $i=1,\ldots, n$.
Let us assume that globally well-defined action-variables
with globally well-defined canonical conjugates exist.
By definition,
this means that there exists a phase space $(\hat M, \hat\omega)$
of the form
\be
\hat M=  \cC_n \times \bR^n = \{ (\hat p,\hat q)\,\vert\, \hat p\in \mathcal{C}_n,\, \hat q\in \bR^n\}
\label{2.1}\ee
with a connected open domain $\mathcal{C}_n\subseteq \bR^n$ and
canonical symplectic form
\be
\hat \omega = \sum_{i=1}^n d\hat q_i \wedge d \hat p_i,
\label{2.2}\ee
which is symplectomorphic to $(M, \omega)$ and permits
identification of the Hamiltonians $h_i$ as functions of the action-variables $\hat p_j$.
More precisely, we assume the existence of a symplectomorphism
\be
A: M\to \hat M
\label{2.3}\ee
such that the functions $ h_i \circ A^{-1}$ do not depend on $\hat q$ and
\be
X_{i,j}:= \frac{\partial h_i\circ A^{-1}}{\partial \hat p_j}
\label{2.4}\ee
yields an invertible matrix $X(\hat p)$ at every  $\hat p\in \mathcal{C}_n$.
As in \cite{AF}, the map $A$ is referred to as a \emph{global action-angle map of maximally
non-compact type}.
The target  $(\hat M,\hat \omega)$ of $A$ is often called the action-angle phase space of the
system $(M, \omega, H)$.

To clarify our conventions, note that
for any real function $F\in C^\infty(\hat M)$ the Hamiltonian vector field ${\boldsymbol{X}}_F$
is here defined by
\be
dF = \hat\omega(\,\cdot\,, \boldsymbol{X}_F),
\label{2.5}\ee
and the Poisson bracket of two functions $F_1, F_2$ reads
\be
\{ F_1,F_2\}_{\hat M} = dF_1(\boldsymbol{X}_{F_2}) = \hat\omega(\boldsymbol{X}_{F_2}, \boldsymbol{X}_{F_1}).
\label{2.6}\ee
In particular, we have
\be
\{ \hat p_j, \hat q_k\}_{\hat M} = \delta_{j,k},
\quad
\{ \hat p_j, \hat p_k\}_{\hat M}= \{ \hat q_j, \hat q_k\}_{\hat M}=0.
\label{2.7}\ee

If a global action-angle map of maximally non-compact type exists,
then one can introduce  functions $f_i \in C^\infty(M)$ ($i=1,\ldots,n)$ by the definition
\be
(f_i \circ A^{-1})(\hat p,\hat q) := \sum_{j=1}^n \hat q_j X(\hat p)^{-1}_{j,i}
\quad
\hbox{with}\quad
\sum_{j=1}^n X(\hat p)_{i,j} X(\hat p)^{-1}_{j,k} = \delta_{i,k}.
\label{2.8}\ee
By using that $A$ is a symplectomorphism, one  obtains the Poisson brackets
\be
\{ h_i, f_j\}_{M} = \delta_{i,j},
\quad
\{ f_i, f_j\}_{M}=0.
\label{2.9}\ee
Indeed, the first relation is immediate from
$
\{ h_i \circ A^{-1}, f_j \circ A^{-1}\}_{\hat M} =
\sum_{k=1}^n \frac{\partial h_i\circ A^{-1}}{\partial \hat p_k}
\frac{\partial f_j\circ A^{-1}}{\partial \hat q_k}
$,
 and the second relation is also easily checked.
Together with $\{ h_i, h_j\}_{M}=0$,  (\ref{2.9})  implies that
the $2n$ functions $h_1,\ldots, h_n, f_1, \ldots, f_n$ are functionally independent
at every point of $M$.

It is plain that
the choice of any of the $2n$ functions $h_1,\ldots, h_n, f_1,\ldots, f_n$
as the Hamiltonian yields a maximally superintegrable system.
For example, the $(2n-1)$ independent functions  $h_1,\ldots, h_n, f_1, \ldots, f_{n-1}$
 Poisson commute with $h_n$.
Under mild conditions, it can be shown \cite{AF} that the generic Liouville integrable Hamiltonian of the form
$H=\cH(h_1,\ldots, h_n)$  is also maximally superintegrable.

\section{Hyperbolic Sutherland and rational RS systems}

We below explain that the hyperbolic Sutherland system
and the rational Ruijsenaars-Schneider  system  admit global action-angle maps of maximally
non-compact type, which implies their maximal superintegrability through the simple construction
presented in the previous section. Remarkably, the pertinent two action angle-maps are the
inverses of each other.

\subsection{Definition of the systems}

The hyperbolic Sutherland system \cite{Sut,CalRag} lives on
the phase space
\be
M:=\mathcal{C}_n\times \bR^n = \{ (q,p)\,\vert\, q\in \mathcal{C}_n,\, p\in \bR^n\,\}
\label{3.1}\ee
with the domain
\be
\mathcal{C}_n=\{ q\in \bR^n \,\vert\, q_1 > q_2 >\cdots > q_n\}.
\label{3.2}\ee
The symplectic form is the canonical one
\be
\omega = \sum_{j=1}^n dp_j \wedge dq_j.
\label{3.3}\ee
A family of $n$ independent commuting Hamiltonians is given by
\be
h_k(q,p):= \tr(L(q,p)^k),\qquad
k=1,\ldots, n,
\label{3.4}\ee
where $L(q,p)$ is the $n\times n$ Hermitian Lax matrix having the entries
\be
L(q,p)_{j,k}:= p_j \delta_{j,k} + \ri (1-\delta_{j,k}) \frac{\kappa}{ \sinh(q_j - q_k)},
\label{3.5}\ee
using a non-zero real parameter $\kappa$.
 The flows of the $h_k$ are complete, and
the main Hamiltonian of interest is
\be
H(q,p):= \frac{1}{2} h_2(q,p) = \frac{1}{2}\sum_{k=1}^n p_k^2 +
\sum_{1\leq j< k\leq n}\frac{\kappa^2}{\sinh^2(q_j-q_k)}.
 \label{3.6}\ee
Thus $q_i$ ($i=1,\ldots, n)$ can be interpreted
as the positions of $n$ interacting particles moving on the line,  restricted
to the domain $\mathcal{C}_n$ by energy conservation.

 The rational Ruijsenaars-Schneider (RS) system \cite{RS} lives on the same phase space, but
 for later purpose we now denote the phase space points as pairs $(\hat p,\hat q)$.
 That is,  the RS phase space is the symplectic manifold $(\hat M, \hat \omega)$
 with\footnote{The notation anticipates that $(\hat M, \hat \omega)$ is the action-angle phase space
 of the Sutherland system $(M,\omega,H)$.}
 \be
 \hat M:=  \mathcal{C}_n\times \bR^n = \{ (\hat p,\hat q)\,\vert\, \hat p\in \mathcal{C}_n,\, \hat q\in \bR^n\,\},
 \qquad
 \hat \omega = \sum_{j=1}^n d \hat q_j \wedge d\hat p_j.
 \label{3.7}\ee
 Now a basic set of Liouville integrable Hamiltonians is provided by  $\hat h_l\in C^\infty(\hat M)$
 for $l=1,\ldots, n$,
 where we define
 \be
 \hat h_l(\hat p, \hat q):= \tr(\hat L(\hat p, \hat q)^l),
 \qquad
 \forall l\in \bZ.
 \label{3.8}\ee
 Here, $\hat L$ is the (positive definite) RS Lax matrix  having the entries
\be
\hat L(\hat p, \hat q)_{j,k}:=
u_j(\hat p,\hat q)
 \left[ \frac{2\ri \kappa}{ 2\ri \kappa + (\hat p_j-\hat p_k)} \right] u_k(\hat p,\hat q),
\label{3.9}\ee
where the $\bR_+$-valued functions $u_j(\hat p, \hat q)$ are given by
\be
u_j(\hat p, \hat q):=  e^{-\hat q_j} z_j(\hat p)^\frac{1}{2}
\quad\hbox{with}\quad
z_j(\hat p):= \prod_{\substack{m = 1 \\ (m \neq j)}}^{n}
 \left[ 1 + \frac{ 4\kappa^2}{(\hat p_j - \hat p_m)^2}\right]^\frac{1}{2}.
\label{3.10} \ee
 In our convention, the principal RS Hamiltonian
 $\hat H = \frac{1}{2}(\hat h_1 + \hat h_{-1})$ reads
 \be
 \hat H(\hat p, \hat q)=
  \sum_{k=1}^n (\cosh 2\hat q_k) \prod_{\substack{j = 1 \\ (j \neq k)}}^{n}\left[ 1+
\frac{4\kappa^2}{(\hat p_k - \hat p_j)^2}\right]^\frac{1}{2},
 \label{3.11}\ee
 and can be viewed as describing $n$ interacting `particles'  with \emph{positions}
 $\hat p_k$ $(k=1,\ldots, n)$.

 \subsection{Dual gauge slices in symplectic reduction}

 Ruijsenaars \cite{SR88} discovered an intriguing duality relation between the pertinent two
 integrable many-body systems, which he called action-angle duality.
 Next we recall the
 geometric interpretation  of this duality, nowadays also called `Ruijsenaars duality',
 following the joint work of  Klim\v c\'\i k with one of us \cite{FK}.

Let $G$ denote the real Lie group $GL(n,\bC)$ and identify the dual space of the
corresponding real Lie algebra $\cG:= gl(n,\bC)$ with itself  using
the invariant bilinear form
\be
\langle X, Y\rangle := \Re \tr(XY),
\qquad
\forall X,Y\in \cG.
\label{3.12}\ee
Consider the minimal coadjoint orbit $\cO_\kappa$ of the group $U(n)$
given as a set by
\be
\cO_{\kappa}:= \{ \ri\kappa (v v^\dagger-\1_n)   \,\vert\, v\in \bC^n,\, \vert v \vert^2 = n\}.
\label{3.13}\ee
Here $v$ is viewed as a column vector, we identified $\fu(n)$ with its dual by the restriction
of the scalar product (\ref{3.12}), and shall also use the notation
\be
\zeta(v):= \ri\kappa (v v^\dagger-\1_n).
\label{3.14}\ee
Trivializing $T^*G$ by means of left-translations,
we introduce the `extended cotangent bundle'
\be
P^\ext:= T^* G \times \cO_\kappa \equiv G \times \cG\times \cO_\kappa  =
\{ (g, J, \zeta)\,\vert\, g\in G,\,\, J \in \cG,\,\,\zeta \in \cO_\kappa\}.
\label{3.15}\ee
 The symplectic form of $P^\ext$ can be
written as
\be
\Omega^\ext = d \langle J, g^{-1} dg \rangle + \Omega^{\cO_\kappa}
\label{3.16}\ee
where  $\Omega^{\cO_\kappa}$ is the standard (Kirillov-Kostant-Souriau) symplectic form of $\cO_\kappa$.

Our basic tool is symplectic reduction of $(P^\ext,\Omega^\ext)$ by the group
\be
K:= U(n) \times U(n)
\label{3.17}\ee
 acting via the symplectomorphisms
\be
\Psi_{\eta_L,\eta_R}(g,J,\zeta) := (\eta_L g \eta_R^{-1}, \eta_R J \eta_R^{-1}, \eta_L \zeta \eta_L^{-1}),
\qquad \forall (\eta_L, \eta_R)\in K.
\label{3.18}\ee
The momentum map $\Phi: P^\ext \to \fu(n) \oplus \fu(n)$ that generates this action is given by
\be
\Phi(g, J,\zeta)= ( (gJ g^{-1})_{\fu(n)} + \zeta, - J_{\fu(n)}),
\label{3.19}\ee
where $X_{\fu(n)}= \frac{1}{2}(X - X^\dagger)$ is the anti-Hermitian part of any $X\in \cG$.
The reduction is defined by setting the momentum map to zero. The associated reduced phase space
\be
P^\red := \Phi^{-1}(0)/K
\label{3.20}\ee
turns out to be a smooth symplectic manifold, with reduced symplectic form $\Omega^\red$.
The point is that the $K$-orbits in the `constraint surface' $\Phi^{-1}(0)$ admit two
global
cross sections that give rise to natural identifications
of the  reduced phase space $(P^\red, \Omega^\red)$ with the
Sutherland phase space $(M,\omega)$ and the RS phase space $(\hat M, \hat \omega)$, respectively.

The first cross section is the so-called `Sutherland gauge slice' $S \subset  \Phi^{-1}(0)$
defined by
\be
S:= \{\, (e^{\bsq}, L(q,p), \zeta(v_0))\,\vert\, (q,p)\in M\, \},
\label{3.21}\ee
where $\bsq:= \diag(q_1,\ldots, q_n)$ and every component of $v_0\in \bC^n$ is equal to $1$.
In fact, $S$ intersects every $K$-orbit in $\Phi^{-1}(0)$ precisely once, and
with the tautological embedding $\iota_S: S \to P^\ext$ it satisfies
\be
\iota_S^* (\Omega^\ext)= \sum_{k=1}^n d p_k \wedge d q_k = \omega.
\label{3.22}\ee
By its very definition (\ref{3.21}), $S$ can be identified with $M$, and the last equation permits  to view
$(M,\omega)$ as a model of the reduced phase space $(P^\red, \Omega^\red)$.

An alternative model of $(P^\red, \Omega^\red)$ is furnished by the following
`Ruijsenaars gauge slice'
\be
\hat S:= \{\, ( \hat L(\hat p, \hat q)^\frac{1}{2},\hat \bsp, \zeta(v(\hat p,\hat q)))\,\vert\,
(\hat p,\hat q)\in \hat M\, \},
\label{3.23}\ee
where $\hat \bsp=\diag(\hat p_1,\ldots, \hat p_n)$ and
$v(\hat p, \hat q)$ is the vector of squared-norm $n$ given by
\be
v(\hat p, \hat q):= \hat L(\hat p, \hat q)^{-\frac{1}{2}} u(\hat p,\hat q),
\label{3.24}\ee
using the Lax matrix $\hat L$ and the vector $u$ introduced in eqs.~(\ref{3.9}-\ref{3.10}).
In fact, $\hat S$ also intersects every $K$-orbit in $\Phi^{-1}(0)$ precisely once, and
with the tautological embedding $\iota_{\hat S}: \hat S \to P^\ext$ it verifies
\be
\iota_{\hat S}^* (\Omega^\ext)= \sum_{k=1}^n d \hat q_k \wedge d \hat p_k = \hat\omega.
\label{3.25}\ee
Thus, identifying $\hat S$ (\ref{3.23}) with $\hat M$, we see that
$(\hat M, \hat \omega)$ also represents  a model of the reduced phase
space $(P^\red, \Omega^\red)$.
It will be clear shortly that
the two gauge slices $S$ and $\hat S$ are dual to each other in the sense that they geometrically
engender Ruijsenaars' action-angle duality between the Sutherland and the RS systems.

The equality (\ref{3.22}) goes back to \cite{KKS}
and equality (\ref{3.25}) was first proved in \cite{FK}.
The proof presented in \cite{FK} uses the `external information'
that the eigenvalues of $\hat L$ form an Abelian Poisson algebra under the Darboux structure $\hat \omega$.
A completely self-contained direct proof of (\ref{3.25})  will be given in the appendix
of  the present communication.

 \subsection{Action-angle duality and superintegrability}

In the previous subsection we described the equivalences
\be
(M,\omega) \longleftrightarrow (S, \iota_S^*(\Omega^\ext))
\longleftrightarrow
(P^\red,\Omega^\red)
\longleftrightarrow
 (\hat S, \iota_{\hat S}^*(\Omega^\ext)) \longleftrightarrow (\hat M,\hat \omega).
\label{3.26}\ee
By composing  the relevant maps, we obtain a symplectomorphism $A: M \to \hat M$,
$A^*(\hat \omega) = \omega$.
It follows easily from the geometric picture that the
map $A$ operates according to the rule
\be
A: (q,p) \mapsto (\hat p, \hat q)
\label{3.27}\ee
characterized the property
\be
 ( \hat L(\hat p, \hat q)^\frac{1}{2},\hat \bsp, \zeta(v(\hat p,\hat q)))
 =
 (\eta(q,p) e^{\bsq} \eta(q,p)^{-1}, \eta(q,p) L(q,p) \eta(q,p)^{-1}, \eta(q,p) \zeta(v_0) \eta(q,p)^{-1}),
\label{3.28}\ee
where $\eta(q,p) \in U(n)$ is uniquely determined up to right-multiplication by a scalar matrix.

Now we are ready to harvest consequences of the above construction.
When doing so, we view $q_i, p_i$ and $\hat p_i, \hat q_i$ as evaluation
functions on $M$ and on $\hat M$, respectively.
The following statements are readily checked:

\begin{itemize}

\item{
First, the particle-positions $\hat p_i$ of the RS system are converted by the map $A$
into  action-variables $\hat p_i \circ A$ of the Sutherland system, and at the same time the canonical momenta
$\hat q_i$ of the RS system  are converted into the corresponding
non-compact `angle-variables' $\hat q_i \circ A$.
This statement holds since $(\hat p_i \circ A)(q,p)$ are just the ordered eigenvalues of the
Sutherland Lax matrix $L(q,p)$. In short, the RS particle-positions and their conjugates
play the roles of  Sutherland action-variables and their conjugates.}

\item{
Second, since the functions $e^{2q_i} \circ A^{-1}$ on $\hat M$ are just the ordered eigenvalues
of the RS Lax matrix $\hat L$, we see that the Sutherland particle-positions $q_i$ are converted by
$A^{-1}$ into action-variables $q_i \circ A^{-1}$ of the RS system, and  the Sutherland
momenta $p_i$ are  converted into the non-compact angle-variables $p_i \circ A^{-1}$
 of the RS system.
That is, the Sutherland particle-positions and their conjugates play the roles
of RS action-variables and their conjugates.}

\item{Third, the maps $A$ and $A^{-1}$ are global action-angle maps of maximally
non-compact type in the sense defined in Section 2.}

\end{itemize}

To verify the third property for the map $A$, one has to consider the commuting
Hamiltonians $h_k$ of equation (\ref{3.4}), which on the action-angle phase space $\hat M$
take the form
\be
(h_k \circ A^{-1})(\hat p, \hat q) = \sum_{l=1}^n \hat p_l^k.
\label{3.29}\ee
It is easily found from the Vandermonde-determinant formula that
\be
\det\left(\frac{\partial h_k \circ A^{-1}}{\partial \hat p_j}\right)=
n! \prod_{1\leq i < j\leq n} (\hat p_j - \hat p_i).
\label{3.30}\ee
This never vanishes on the domain $\mathcal{C}_n$,  proving the claim.
As for $A^{-1}$, notice from (\ref{3.8}) and (\ref{3.28})  that
\be
(\hat h_k \circ A)(q,p) = \sum_{l=1}^n e^{2k q_l},
\qquad \forall k=1,\ldots, n.
\label{3.32}\ee
It follows that
\be
\det \left( \frac{\partial \hat h_k \circ A}{\partial q_j}\right) = 2^n n! \prod_{k=1}^n e^{2q_k}
\prod_{1\leq i < j \leq n} (e^{2q_j} - e^{2q_i}),
\label{3.31}\ee
and this expression is non-zero  for every $q\in \mathcal{C}_n$.

The fact that $A: M \to \hat M$ is an action-angle map for the Sutherland system
$(M,\omega,H)$  and $A^{-1}: \hat M \to M$ is an action-angle map for the RS system
$(\hat M,\hat \omega, \hat H)$ is  expressed by saying that these two
 many-body systems enjoy `action-angle duality' relation \cite{SR88}.
In particular, each lives on the action-angle phase space of the other
and the position-variables of any of the two systems become action-variables of the other system
under the action-angle map.

The general argument of Section 2 now implies directly
that any of the commuting Hamiltonians $h_1,\ldots, h_n$, and in particular
 the Sutherland Hamiltonian $H=\frac{1}{2} h_2$,  is maximally superintegrable.
 Similarly, any of the commuting Hamiltonians $\hat h_k$ $(k=1,\ldots, n)$
of the RS system is maximally superintegrable.
The principal RS Hamiltonian $\hat H = \frac{1}{2}(\hat h_1 + \hat h_{-1})$
can be expressed as a polynomial in terms of $\hat h_1,\ldots, \hat h_n$, and one
can use this to establish its superintegrability as well \cite{AF}.

At first sight the above reasoning  is independent of scattering theory
that also could be used to establish maximal superintegrability
of the repulsive interactions encoded by $H$ (\ref{3.6}) and $\hat H$ (\ref{3.11}).
This is somewhat an illusion, however, since the  action-angle maps
$A$ and $A^{-1}$ are  closely related to the scattering wave maps
of the systems under consideration \cite{SR88}.
Nevertheless, an advantage of our arguments is that they
do not require any analysis of
the large time asymptotic of the dynamics, which is needed in scattering theory.
Instead, our reasoning is based  on the elegant geometry of
the underlying symplectic reduction.

\subsection{Explicit extra constants of motion in the  RS system}

The key equation (\ref{3.28}) leads to an algebraic algorithm  for constructing
the maps $A$  and $A^{-1}$ in terms of diagonalization of the Lax matrices $L$ and $\hat L$.
However, explicit formulae of these action-angle maps are not available.
Thus non-trivial effort is required to find extra constants of motion in explicit form
both for the rational RS and for the hyperbolic Sutherland system.
In the former case, this problem was solved in \cite{AF}.

The work reported in \cite{AF} was inspired by Wojciechowski's paper \cite{W}
that explicitly established the superintegrability
of the rational Calogero Hamiltonian.
In the RS case,  since the Lax matrix $\hat L$ (\ref{3.9}) is positive definite, one can
define the  smooth real functions
\be
 \hat h_j(\hat p,\hat q):= \tr\left( \hat L(\hat p,\hat q)^j\right),
    \quad
    \hat h_k^1(\hat p,\hat q):= \tr\left(  \hat L(\hat p,\hat q)^k \boldsymbol{\hat p} \right),
    \quad \forall j, k\in\bZ.
 \label{3.33}\ee
It turned out that these functions satisfy the following Poisson algebra:
\be
\{ \hat h_k, \hat h_j\}_{\hat M}= 0, \quad \{ \hat h_k^1, \hat h_j\}_{\hat M} = j \hat h_{j+k}, \quad
 \{ \hat h_k^1, \hat h_j^1\}_{\hat M}= (j-k) \hat h_{k+j}^1.
\label{3.34}\ee
The relations (\ref{3.34}) were proved in \cite{AF}  utilizing
the symplectic reduction described in Subsection 3.2.

The basic reason for which the (first two) relations of (\ref{3.34}) are useful
in investigating superintegrability is as follows.
Take an \emph{arbitrary} Liouville integrable Hamiltonian
\be
\hat H=\cH(\hat h_1,\ldots, \hat h_n).
\label{3.35}\ee
Observe that this Hamiltonian Poisson commutes not only with all the $\hat h_j$, but also
with all functions of the form
\be
C_{j,k}^{\hat H}:= \hat h_k^1 \{\hat h_j^1, \hat H\}_{\hat M} - \hat h_j^1\{\hat h_k^1, \hat H \}_{\hat M},
\qquad
\forall j,k\in \bZ.
\label{3.36}\ee
Then one should select $(n-1)$ functions out of this set so that together
with $\hat h_1,\ldots, \hat h_n$ they imply the maximal superintegrability of $\hat H$.
To show functional independence, the selection must use the concrete form of the functions that appear.

As a special case, it was found in \cite{AF} that
for any fixed $j\in \{1,\ldots, n\}$ the functions
\be
 C_{j,k}^{\hat h_j} = j \hat h_k^1 \hat h_{2j} -  j \hat h_j^1 \hat h_{j+k},
\qquad
k\in \{1,\ldots,n\} \setminus \{j\}
\label{3.37}\ee
that commute with $\hat h_j$
form an independent set together with $\hat h_1,\ldots, \hat h_n$.
Furthermore, a set of `extra constants of motion' that explicitly shows
the superintegrability of the RS Hamiltonian $\hat H=\frac{1}{2}( \hat h_1 + \hat h_{-1})$  is provided by
 \be
 \hat F_j:=\hat h_j^1 ( \hat h_2 - n) - \hat h_{1}^1 (\hat h_{j+1} - \hat h_{j-1}), \qquad
 j=2,\ldots,n.
\label{3.38}\ee

It is worth noting that the quantities $\hat h_k^1$ are useful not only for constructing the
constants of motion (\ref{3.36}), but also since their time development along the
solutions $x(t)=(\hat p(t),\hat q(t))$ of the system $(\hat M,\hat \omega, \hat H)$,
for any Hamiltonian (\ref{3.35}), is especially  simple.
Namely, since $\{\{\hat h_k^1,\hat H\}, \hat H\}_{\hat M}=0$ follows from (\ref{3.34}),
we obtain that
\be
\hat h_k^1(x(t)) = \hat h_k^1(x(0)) + t \{h_k^1, \hat H \}_{\hat M}(x(0))
\label{3.39}\ee
is linear in time.
In this way, $\hat h_k$ and $\hat h_k^1$ ($k=1,\ldots, n$) linearize the dynamics.
This is similar to the linearization provided by the non-compact analogues
of action-angle variables, with the distinctive feature that
 $\hat h_k$ and $\hat h_k^1$ are
\emph{explicitly} given functions on the phase space.

\section{Conclusion}

In this paper we explained that the hyperbolic Sutherland and the rational
RS systems are both maximally superintegrable since Ruijsenaars'
duality symplectomorphism \cite{SR88} between these two systems qualifies as a global action-angle
map of maximally non-compact type, and every Liouville integrable system that
possesses such action-angle map is maximally superintegrable.
Although these results are certainly known to experts, we
hope that our self-contained
exposition based on the geometric interpretation of the duality \cite{FK}
may be useful, especially since it can be applied to other examples as well.

Indeed, essentially the same arguments can be applied to the
$BC_n$ generalizations of the Sutherland and RS systems, which are encoded by the
 Hamiltonians
\bea
&& H_{BC}(q,p) =  \frac{1}{2}\sum_{c = 1}^{n} p_c^2
+
\sum_{1\leq a < b\leq n} \left( \frac{g^2}{\sinh^2(q_a-q_b)} + \frac{g^2}{\sinh^2(q_a+q_b)}
\right)
\nonumber \\
&& \phantom{ H_{BC}(q,p) =}
 +  \sum_{c = 1}^n  \left(\frac{g_1^2}{\sinh^2 q_c} + \frac{g_2^2}{\sinh^2 (2q_c)}\right)
\label{4.1}\eea
and
\bea
&& \hat H_{BC}(\hat p,\hat q) =  \sum_{c = 1}^{n} (\cosh 2\hat q_c)
\left[1 + \frac{\nu^2} {\hat p_c^2}\right]^\frac{1}{2}
\left[1 + \frac{\chi^2}{\hat p_c^2}\right]^\frac{1}{2}
\prod_{\substack{d = 1 \\ (d \neq c)}}^{n}
\left[ 1 + \frac{ 4\mu^2}{(\hat p_c - \hat p_d)^2} \right]^\frac{1}{2}
\left[1+ \frac{ 4\mu^2} {(\hat p_c + \hat p_d)^2} \right]^\frac{1}{2}
\nonumber \\
&& \phantom{\hat H_{BC}(\hat p,\hat q) =}
+ \frac{\nu \chi}{ 4\mu^2}\prod_{c = 1}^n \left(1 + \frac{ 4\mu^2}{\hat p_c^2}\right)
- \frac{\nu \chi}{ 4\mu^2}.
\label{4.2}\eea
The $BC_n$ Sutherland system (\ref{4.1}) was introduced by Olshanetsky and Perelomov \cite{OP},
while the $BC_n$ variant of the RS system (\ref{4.2}) is largely due to
van Diejen \cite{vD}.
In a recent work \cite{P2},  Pusztai proved by using a suitable symplectic reduction that these two
systems are in action-angle duality if their respective 3 coupling parameters are
related according to
\be
g^2 = \mu^2,
\qquad
g_1^2 = \frac{1}{2} \nu \chi,
\qquad
g_2^2= \frac{1}{2}(\nu -\chi)^2
\ee
with arbitrary $\mu^2 >0$, $\nu>0$ and $\chi \geq 0$.
The duality symplectomorphism is again given by the natural map
between two gauge slices, and it yields action-angle maps of maximally non-compact
type analogously to the $A_{n-1}$ case.

Finally, we remark that $BC_n$ analogues of the extra constants of motion
presented in Subsection 3.4 are still not known,
so it  could be worthwhile to search for such constants of motion, and to search also
for explicit constants of motion in the hyperbolic Sutherland systems.

\renewcommand{\thesection}{\Alph{section}}
\setcounter{section}{0}

\bigskip
\section{Reduced symplectic form in the Ruijsenaars gauge}

The goal of this appendix is to give a self-contained proof of  formula (\ref{3.25}),
which describes the reduced symplectic structure in terms of the Ruijsenaars gauge slice $\hat S$ (\ref{3.23}).
A rather roundabout proof was presented in \cite{FK}.
Here, we adopt the method  of Pusztai \cite{P1}.

We identify the reduced phase space $P^\red$ (\ref{3.20}) with the global gauge slice $\hat S$,
whereby the reduced symplectic form becomes
\be
\Omega^\red \equiv \iota_{\hat S}^*(\Omega^\ext).
\label{A1}\ee
Then, by means of the parametrization of $\hat S$ in (\ref{3.23}),
 we regard the components of $\hat p$ and $\hat q$ as coordinates on $\hat S$.
Let us denote the Poisson bracket of arbitrary functions $F_1^\red, F_2^\red \in C^\infty(\hat S)$
 determined by means on $\Omega^\red$ as $\{ F_1^\red, F_2^\red\}$.
We wish to find the Poisson brackets
\be
\{ \hat p_\alpha, \hat p_\beta\},
\quad
\{ \hat p_\alpha, \hat q_\beta\},
\quad
\{ \hat q_\alpha, \hat q_\beta\}.
\label{A2}\ee

As a preparation, we introduce the following functions
$\varphi_m, \psi_k \in C^\infty(P^\ext)^K$,
\be
\varphi_m(g,J,\zeta)=\frac{1}{2m}\tr(J^m+(J^\dag)^m),
\qquad
\psi_k(g,J,\zeta)=\frac{1}{2}\tr((J^k+(J^\dag)^k)g^\dag Z(\zeta)g),
\label{A3}\ee
where $m\geq 1$, $k\geq 0$ are integers and
\be
Z(\zeta):=(\ri\kappa)^{-1}\zeta+\1_n,
\qquad
\forall \zeta\in \cO_\kappa.
\label{A4}\ee
It is easily seen that these functions are indeed invariant under the $K$-action (\ref{3.18}).
We also consider the corresponding reduced functions
\be
\varphi_m^\red:= \iota_{\hat S}^*(\varphi_m),
\qquad
\psi_k^\red:= \iota_{\hat S}^*(\psi_k).
\label{A5}\ee
These functions belong to $C^\infty(\hat S)$ and have the form
\be
\varphi_m^\red(\hat{p},\hat q)=\frac{1}{m}\sum_{j=1}^n \hat{p}_j^m,
\qquad
\psi_k^\red(\hat{p},\hat{q})=\sum_{j=1}^n \hat{p}_j^k z_j(\hat{p}) e^{-2\hat{q}_j},
\label{A6}\ee
with the vector $z(\hat p)$ defined in (\ref{3.10}).

If
$F_i^\red = \iota_{\hat S}^*(F_i)$ for some
$F_i \in C^\infty(P^\ext)^K$ ($i=1,2$),
then the definition of symplectic reduction implies
\be
\iota_{\hat S}^* (\{ F_1, F_2\}^\ext) = \{ F_1^\red, F_2^\red\},
\label{A7}\ee
where the Poisson bracket on the left-hand-side is computed on $(P^\ext, \Omega^\ext)$.
The idea is to extract the required Poisson brackets in (\ref{A2}) from equality (\ref{A7})
applied to various
choices of $F_1, F_2$ from the set of functions $\varphi_m, \psi_k$.
Note that $\{ F_1, F_2\}^\ext = \Omega^{\ext}(\bX_{F_2}, \bX_{F_1})$
with the corresponding Hamiltonian vector fields.

An arbitrary vector field $\bX$ on $P^\ext$ (\ref{3.15}) can be written as
$\bX = \Delta g\oplus\Delta J\oplus\Delta\zeta$, where at $(g,J,\zeta)\in P^\ext$ one has
 $\Delta g \in T_gG$, $\Delta J \in T_J \cG\simeq \cG$ and $\Delta \zeta \in T_\zeta \cO_\kappa$.
Evaluation of the symplectic form (\ref{3.16}) on  two vector fields $\bX$ and $\bX'$
  yields the function
 \be
 \Omega^\ext(\bX, \bX')= \langle g^{-1} \Delta' g, \Delta J \rangle - \langle g^{-1} \Delta g, \Delta' J \rangle
 + \langle [g^{-1} \Delta' g, g^{-1} \Delta g], J\rangle - \langle \zeta, [D_\zeta, D_\zeta']\rangle,
 \label{A8}\ee
 where in the last term we use $\Delta \zeta = [D_\zeta, \zeta]$ and $\Delta' \zeta = [D'_\zeta, \zeta]$
 with some $\frak{u}(n)$-valued $D_\zeta$ and $D'_\zeta$.
It is not difficult to verify the following formulae of the
Hamiltonian vector fields of $\varphi_m$ and $\psi_k$:
\be
\bX_{\varphi_m}=gJ^{m-1}\oplus 0\oplus 0\quad
\hbox{and}\quad
\bX_{\psi_k}=\Delta g\oplus\Delta J\oplus\Delta\zeta
\label{A9}
\ee
with components
\begin{eqnarray}
&&\Delta g= g\sum_{j=0}^{k-1}J^jg^\dag Z(\zeta)gJ^{k-1-j},
\label{A10}\\
&&\Delta J=-(J^\dag)^k g^\dag Z(\zeta)g-g^\dag Z(\zeta)g J^k,
\label{A11}\\
&&\Delta\zeta=\frac{1}{2\ri\kappa}[g(J^k+(J^\dag)^k)g^\dag,\zeta].
\label{A12}
\end{eqnarray}
Note that for $k=0$ the sum in (\ref{A10}) is vacuous and in this special case $\Delta g=0$.

\medskip
\noindent
{\bf Lemma 1.}
\emph{We have $\{\hat{p}_\alpha,\hat{p}_\beta\}=0$ for all $\alpha,\beta=1,\ldots, n$.}
\medskip

\noindent
{\bf Proof.}
We readily derive from the above that $\{\varphi_m,\varphi_l\}^\ext =0$
for any $m,l\in\bN$, which immediately results in
$\{\varphi_m^\red,\varphi_l^\red\}=0$.
On the other hand, using only
the basic properties of the Poisson bracket such as bilinearity and Leibniz rule, we obtain from
the formula (\ref{A6}) of these functions that
\be
\{\varphi_m^\red,\varphi_l^\red\}=
\sum_{\alpha,\beta=1}^n \hat{p}_\alpha^{m-1}\{\hat{p}_\alpha,\hat{p}_\beta\}\hat{p}_\beta^{l-1}.
\label{A13}\ee
Now let us introduce the $n\times n$ matrices
$ \mathcal{P}_{\alpha,\beta}:=\{\hat{p}_\alpha,\hat{p}_\beta\}$
and
\be
V_{\alpha,\beta}:=\hat{p}_\alpha^{\beta-1}, \qquad \alpha,\beta=1,2,\ldots,n.
\label{A14}\ee
Notice that $V$ is a Vandermonde matrix and its
determinant is non-zero (as $\hat{p}_1>\hat{p}_2>\dots>\hat{p}_n$).
Taking $m, l$ from the set $\{1,\ldots, n\}$,
 we can write (\ref{A14}) in matrix form
\be
\sum_{\alpha,\beta=1}^n \hat{p}_\alpha^{m-1}
\{\hat{p}_\alpha,\hat{p}_\beta\}\hat{p}_\beta^{l-1}=\sum_{\alpha,\beta=1}^n V_{\alpha,m}
\mathcal{P}_{\alpha,\beta}V_{\beta,l}=(V^\dag \mathcal{P} V)_{m,l}.
\label{A15}\ee
 Because this expression must vanish and $V$ is invertible, it follows that $\mathcal{P}=0$,
 i.e., $\{\hat{p}_\alpha,\hat{p}_\beta\}=0$ for all $\alpha, \beta =1,\ldots, n$.
\hspace*{\stretch{1}} \qed

\medskip
\noindent
{\bf Lemma 2.} \emph{We have $\{\hat{p}_\alpha,\hat{q}_\beta\}=\delta_{\alpha, \beta}$
for all $\alpha,\beta=1,\ldots, n$.}

\medskip
\noindent
{\bf Proof.}
Taking arbitrary
 \be
  k=0,1,\ldots, n-1 \quad\hbox{and}\quad
 l =1,\ldots, n,
 \label{extra1}\ee
  it can be checked that
$\{\psi_k,\varphi_l\}^\ext =2\psi_{k+l-1}$ holds at all triples $(g,J,\zeta)$ for which
$J=J^\dagger$.
 Hence we must have
\be
\{\psi_k^\red,\varphi_l^\red\}=2\psi_{k+l-1}^\red.
\label{A16}\ee
Using the basic properties of the Poisson bracket and the statement of Lemma 1,
we can directly calculate this Poisson bracket as
\be
\{\psi_k^\red,\varphi_l^\red\}=
\sum_{\alpha=1}^n\hat{p}_\alpha^k z_\alpha e^{-2\hat{q}_\alpha}
\sum_{\beta=1}^n\{-2\hat{q}_\alpha,\hat{p}_\beta\} \hat{p}_\beta^{l-1}.
\label{A17}\ee
The comparison of the last two equations leads to
\be
\sum_{\alpha=1}^n\hat{p}_\alpha^k z_\alpha e^{-2\hat{q}_\alpha}
\bigg(\sum_{\beta=1}^n \{-2\hat{q}_\alpha,\hat{p}_\beta\}
\hat{p}_\beta^{l-1}-2\hat{p}_\alpha^{l-1}\bigg)=0.
\label{A18}\ee
By introducing the $n\times n$ matrix
\be
\mathcal{A}_{\alpha,\beta}=z_\alpha e^{-2\hat{q}_\alpha}
\bigg(\sum_{\gamma=1}^n \{-2\hat{q}_\alpha,\hat{p}_\gamma\}\hat{p}_\gamma^{\beta-1}-
2\hat{p}_\alpha^{\beta-1}\bigg)
\label{A19}\ee
we can write (\ref{A19}) as
$(V^\dag \mathcal{A})_{k+1,l}=0$. Since $V$ (\ref{A14}) is invertible,  we conclude that $\mathcal{A}=0$.
   Now if we collect the expressions $\{-2\hat{q}_\alpha,\hat{p}_\beta\}$ in the $n\times n$ matrix
$\mathcal{B}_{\alpha,\beta}:=\{-2\hat{q}_\alpha,\hat{p}_\beta\}$,
then the vanishing of $\mathcal{A}$ can be re-stated as the matrix equation
$\mathcal{B} V-2V=0$.
This entails that $\mathcal{B}=2\1_n$,
which is  equivalent to
 $\{\hat{p}_\alpha,\hat{q}_\beta\}=\delta_{\alpha,\beta}$
for all $\alpha,\beta$.
\hspace*{\stretch{1}} \qed

\medskip
\noindent
{\bf Lemma 3.} \emph{We have $\{\hat{q}_\alpha,\hat{q}_\beta\}=0$ for all $\alpha, \beta=1,\ldots, n$.}

\medskip
\noindent
{\bf Proof.}
We now determine the reduced Poisson bracket
\be
\{\psi_k^\red,\psi_l^\red\},\qquad
\forall
k,l= 0,1,\ldots, n-1,
\label{extra2}\ee
 in two ways. First we use
$\{\psi_k^\red,\psi_l^\red\} =\{\psi_k,\psi_l\}^\ext\circ \iota_{\hat S}$ and obtain
by calculating the right-hand-side that
\be
\{\psi_k^\red,\psi_l^\red\}=
-2(k-l)\sum_{\alpha=1}^n\hat{p}_\alpha^{k+l-1}z_\alpha^2e^{-4\hat{q}_\alpha}
-16\kappa^2\sum_{\substack{\alpha,\beta=1\\(\alpha\neq\beta)}}^n
\frac{\hat{p}_\alpha^k\hat{p}_\beta^l z_\alpha
 z_\beta e^{-2(\hat{q}_\alpha+\hat{q}_\beta)}}
{(4\kappa^2+(\hat{p}_\alpha-\hat{p}_\beta)^2)(\hat{p}_\alpha-\hat{p}_\beta)}.
\label{20}\ee
Then direct calculation of $\{\psi_k^\red,\psi_l^\red\}$,
utilizing basic properties of the Poisson bracket together with the preceding lemmas,
gives
\bea
&&\{\psi_k^\red,\psi_l^\red\}=2\sum_{\alpha,\beta=1}^n \bigg[\hat{p}_\alpha^k
 z_\alpha \frac{\partial \hat{p}_\beta^l z_\beta }{\partial \hat{p}_\alpha}-
\hat{p}_\beta^k z_\beta \frac{\partial \hat{p}_\alpha^k z_\alpha }{\partial \hat{p}_\beta}\bigg]
e^{-2(\hat{q}_\alpha+\hat{q}_\beta)}\nonumber\\
&&\phantom{\{\psi_k^\red,\psi_l^\red\}=}
+4\sum_{\alpha,\beta=1}^n \hat{p}_\alpha^k\hat{p}_\beta^l z_\alpha z_\beta
e^{-2(\hat{q}_\alpha+\hat{q}_\beta)}\{\hat{q}_\alpha,\hat{q}_\beta\}.
\label{21}\eea
Simple algebraic manipulations permit to spell this out more explicitly
\bea
&&\{\psi_k^\red,\psi_l^\red\}=
-2(k-l)\sum_{\alpha=1}^n\hat{p}_\alpha^{k+l-1} z_\alpha ^2e^{-4\hat{q}_\alpha}
-16\kappa^2\sum_{\substack{\alpha,\beta=
1\\(\alpha\neq\beta)}}^n\frac{\hat{p}_\alpha^k\hat{p}_\beta^l z_\alpha
 z_\beta e^{-2(\hat{q}_\alpha+\hat{q}_\beta)}}
{(4\kappa^2+(\hat{p}_\alpha-\hat{p}_\beta)^2)(\hat{p}_\alpha-\hat{p}_\beta)}\nonumber\\
&&\phantom{\{\psi_k^\red,\psi_l^\red\}=}
+4\sum_{\alpha,\beta=1}^n \hat{p}_\alpha^k\hat{p}_\beta^l  z_\alpha
 z_\beta e^{-2(\hat{q}_\alpha+\hat{q}_\beta)}\{\hat{q}_\alpha,\hat{q}_\beta\}.
\label{22}\eea
Comparing equations (\ref{20}) and (\ref{22}), we then find that
\be
\sum_{\alpha,\beta=1}^n\hat{p}_\alpha^k\hat{p}_\beta^l z_\alpha z_\beta
e^{-2(\hat{q}_\alpha+\hat{q}_\beta)}\{\hat{q}_\alpha,\hat{q}_\beta\}=0.
\label{A23}
\ee
Inspecting this equation using the non-degeneracy of the matrix $V$ (\ref{A14})
and that the functions $z_\alpha$ never vanish, we find that
$\{\hat{q}_\alpha,\hat{q}_\beta\}$ must vanish for all $\alpha$ and $\beta$.
\hspace*{\stretch{1}} \qed

\bigskip
The three lemmas together prove the important formula (\ref{3.25}), which was proved in \cite{FK} by
a less self-contained  method.

\medskip
\bigskip
\bigskip
\noindent{\bf Acknowledgements.}
Support by the Hungarian Scientific Research Fund under the grant OTKA K77400 is
hereby acknowledged.
This publication was also supported by the European Social
Fund under the project number T\'AMOP-4.2.2/B-10/1-2010-0012.

\end{document}